\documentclass[showpacs,preprintnumbers,amsmath,amssymb,pre]{revtex4}

\usepackage{epsfig,graphics,graphicx}

\begin{document}

\title{Exact solution of a stochastic protein dynamics model with delayed degradation}
\author{L. F. Lafuerza, R. Toral}

\affiliation{IFISC, Instituto de F{\'\i}sica Interdisciplinar y Sistemas Complejos, CSIC-UIB,  Campus UIB, E-07122 Palma de Mallorca, Spain}

\date{\today}

\pacs{02.50.Ey, 05.10.Gg, 05-40.a, 87.18.Tt.}

\begin{abstract}
We study a stochastic model of protein dynamics that explicitly includes delay in the degradation. We rigorously derive the master equation for the processes and solve it exactly. We show that the equations for the mean values obtained differ from others intuitively proposed and that oscillatory behavior is not possible in this system. 
We discuss the calculation of correlation functions in stochastic systems with delay, stressing the differences with Markovian processes. The exact results allow to clarify the interplay between stochasticity and delay.
\end{abstract}

\maketitle
\section{Introduction}
Due to the small number of some molecules involved and to the uncontrolled environment, biochemical processes inside a cell usually need to be described by stochastic models\cite{elowitz,oudenaarden}. Some important basic processes (such as transcription, translation or specific degradation) are indeed compound multistage reactions involving a large number of steps of similar duration, and, in principle, due to the central limit theorem, the time to complete such processes should follow nearly a Gaussian distribution, rather than exponential, with a well defined characteristic delay time \cite{morelli}. A description including delay is then needed to obtain a reduced model for this kind of processes.
It is well known that delay can change qualitatively the dynamical behavior, allowing, for example, the appearance of oscillations in the evolution of the number of molecules\cite{makeyglass}, and there has been a great interest in delay-induced oscillations in biological systems.

Stochastic processes that include delay are analytically difficult due to their non-Markovian character. Most theoretical studies consider a Langevin approach (stochastic differential equations)\cite{langevindelay,Frank} or systems in discrete time\cite{rwdelay} (where delay can be accounted for by increasing the number of variables). None of these approaches is completely suitable to describe chemical reactions inside a cell since the former neglects the inherently discrete nature of the molecule levels and the latter considers an arbitrary discretization of time. In some cases, the discreteness can be a major source of fluctuations\cite{aparicio}. 

In this work we develop a rigorous derivation of the stochastic description of a protein dynamics model that includes delay in the degradation, and solve it exactly. We find that the exact solution for the probabilities leads to equations for the mean values that do not comply with simple intuitive arguments and that oscillatory behavior does not exist (while it is usually believed to be present in this type of system). This clarifies and warns about the derivation of dynamical equations describing the evolution of the concentrations in cases in which delay plays a role. The exact solution is specially valuable for small system sizes, where approximated schemes typically fail. Due to the low number of molecules inside cells, this regime may be biologically relevant. Our solution extends the results of reference \cite{Mieckisz} in which the authors consider a particular case which allows for a Markovian description in terms of suitably defined dynamical variables, while such a Markovian reduction has not been achieved in the more complete case studied here.

The paper is organized as follows: in the next section we define the model and in section \ref{master} we derive the corresponding master equation. In section \ref{exactrates} we solve the master equation and derive the probabilities and the (macroscopic) equations for the mean values in the case of constant rates. Some technical details are left for the Appendix \ref{appendix}. In section \ref{correlations} we discuss the details of the derivation of the master equation for the conditional probabilities and derive, again in the case of constant rates, the correlation functions of the process. Finally, in section \ref{conclusions} we end with a brief discussion of the results.

\section{Model}\label{model}
Proteins usually degrade through complex proteolytic pathways that involve several different steps (such as tagging, binding of auxiliary proteins, recognition by protease and destruction)\cite{levchenko}. As discussed before, it is then natural to consider a time delay in this process. As a simple model for the dynamics of the number of molecules of a protein $X$ we consider:
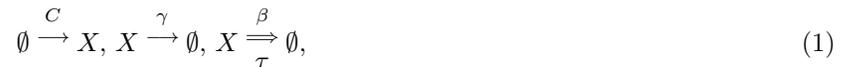
\begin{equation}
\emptyset {{C \atop \longrightarrow}\atop{}} X, \, X {{\gamma \atop \longrightarrow}\atop{}} \emptyset, \,X {{{\beta} \atop \Longrightarrow}\atop{\tau}} \emptyset,
\end{equation}
where, for simplicity and to isolate the effect of delay in degradation, transcription and translation steps have been lumped into a single stochastic process that occurs at a rate $C$.
The delayed degradation (indicated by a double arrow) is modeled as a reaction that is initiated at a rate ${\beta}$ and completed at a time $\tau$ after initiated (giving rise to the destruction of the protein). Besides delayed degradation, we consider also instantaneous degradation at a rate $\gamma$ (which can take into account processes such as non-selective degradation, proteins going to the membrane or outside the cell, dilution due to cell growth, etc.). This model of protein dynamics was proposed in \cite{bratsun} and it was thought that it can lead to periodic oscillations, although the analysis of \cite{bodnar,Mieckisz} in a particular limit showed otherwise. To completely describe the process, one has to specify if a protein that initiates delayed-degradation at time $t$ (and thus will disappear at $t+\tau$) can also disappear before the completion of this reaction, through instantaneous degradation at a rate $\gamma'$, not necessarily equal to $\gamma$.
The process is equivalent to the following two-variable system:
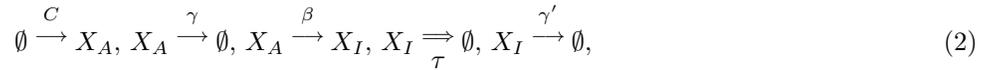
\begin{equation}\label{stoproc}
\emptyset {{C \atop \longrightarrow}\atop{}} X_{A}, \,X_{A} {{\gamma \atop \longrightarrow}\atop{}} \emptyset, \,X_{A} {{{\beta} \atop \longrightarrow}\atop{}} X_{I}, \,X_{I} {{\atop \Longrightarrow}\atop{\tau}} \emptyset, \,X_{I} {{\gamma' \atop \longrightarrow}\atop{}} \emptyset,
\end{equation}
where we have split the proteins into two types: $X_{I}$ are ``infected'' molecules that will die precisely at a time $\tau$ after being infected (if still present) and $X_{A}$ are non-infected particles (so $X=X_{A}\cup X_{I}$). In principle, we allow the rates to depend on $n_A$, the number of $X_A$ particles, but not on  $n_I$, the number of $X_I$ particles which are considered to be ``inert". One has to include an extra variable because delayed degradation is a ``consuming reaction''\cite{barrio}, since as soon as one particle initiates this reaction the number of particles that can initiate it decreases by one, and so the state of the system changes both when the reaction is initiated and when it is finished, after a time $\tau$. In a recent work \cite{Mieckisz}, the stochastic description was studied in the particular case that infected proteins can not undergo instantaneous degradation, $\gamma'=0$. In turns out that this particular case allows for a Markovian description in terms of suitably defined dynamical variables. Our treatment allows us to pose and solve a non-Markovian problem and it can be of interest to other processes not allowable to a Markovian description.

\section{Master equation}\label{master}
To derive the master equation of the process, we start with the following identity, valid for any stochastic process (with a numerable set of states, otherwise the sum should be replaced by an integral):
\begin{equation}
P(n,t+\Delta t)= \sum_{n'}P(n,t+\Delta t;n',t)\label{generaldelayme},
\end{equation}
where $P(n,t)$ is the probability of being at state $n$ at time $t$ and $P(n,t;n',t')$ the joint probability of being at state $n$ at time $t$ and at state $n'$ at time $t'$.
In most cases of interest, $P(n,t+\Delta t;n',t)$ is  $O(\Delta t^{0})$ or $O(\Delta t)$ only for a small number of $n'$ (usually depending on $n$), whereas it is $o(\Delta t)$ for all the rest. Writing explicitly the terms $O(\Delta t^{0})$ and $O(\Delta t)$, dividing by $\Delta t$ and taking the limit $\Delta t\rightarrow0$ one can obtain the master equation of the process (differential equation for $P(n,t)$). For Markovian processes one can write the joint probabilities as $P(n,t+\Delta t;n',t)=P(n',t)P(n,t+\Delta t|n',t)$ and the conditional probabilities can be readily derived from the rules of the process, obtaining in this way a closed equation for the one-time probability $P(n,t)$. For non-Markovian ones, in general, the equation for the one-time probability will depend on higher order joint probability densities $P(n,t;n_{1},t_{1};...;n_{m},t_{m})$. In some cases (as in the one considered here) it is possible to explicitly write the joint probability densities that appear in the master equation and obtain a closed equation for $P(n,t)$.

In our process, $n$ has two components ($n_A$ and $n_I$) and the only terms $O(\Delta t^{0})$ and $O(\Delta t)$ that appear in (\ref{generaldelayme}) are associated to the following elementary processes schematized in (\ref{stoproc}):\\
(1) Birth of an $X_A$ particle from the reservoir: $(n_A-1,n_I)\to (n_A,n_I)$, with probability $C\Delta t$. The contribution to the right-hand-side (rhs) of Eq.(\ref{generaldelayme}) is $P(n_A-1,n_I,t)C\Delta t$.\\
(2) Death of an $X_A$ particle: $(n_A+1,n_I)\to (n_A,n_I)$, with probability $\gamma(n_A+1)\Delta t$. The contribution to the rhs of Eq.(\ref{generaldelayme}) is $P(n_A+1,n_I,t)\gamma(n_A+1)\Delta t$.\\
(3) Infection  of an $X_A$ particle: $(n_A+1,n_I-1)\to (n_A,n_I)$, with probability ${\beta}(n_A+1)\Delta t$. The contribution to the rhs of Eq.(\ref{generaldelayme}) is $P(n_A+1,n_I-1,t){\beta}(n_A+1)\Delta t$.\\
(4) Death of an $X_I$ particle: $(n_A,n_I+1)\to (n_A,n_I)$. This might occur by two different reasons:\\
\indent(4a) Death by instantaneous decay at rate $\gamma'$ with probability $\gamma'(n_I+1)\Delta t$. The contribution to the rhs of Eq.(\ref{generaldelayme}) is $P(n_A,n_I+1,t)\gamma'(n_I+1)\Delta t$.\\
\indent(4b) Death after time $\tau$ after infection. This is the only non-Markovian contribution. In order to account for this case, and according to the previous discussion, we need to write the corresponding contribution to the rhs of Eq.(\ref{generaldelayme}) as:
\begin{eqnarray}
\sum_{n'_A,n'_I}P(n_A,n_I,t+\Delta t;n_A,n_I+1,t;{\cal S};{\cal I}_{n_A',n_I'}),
\end{eqnarray}
where ${\cal I}_{n_A',n_I'}=\{n'_A-1,n'_I+1,t-\tau+\Delta t;n_A',n_I',t-\tau\}$ denotes the event in which there were $n'_A$ and $n'_I$ particles at time $t-\tau$ and one particle got infected during the time interval $(t-\tau;t-\tau+\Delta t)$, and $\cal S$ denotes that the particle infected during that time interval survived up to time $t$. We use now the series of conditional probabilities:
\begin{eqnarray}
P(n_A,n_I,t+\Delta t;n_A,n_I+1,t;{\cal S};{\cal I}_{n_A',n_I'})&=&P(n_A,n_I,t+\Delta t|n_A,n_I+1,t;{\cal S};{\cal I}_{n_A',n_I'})\\
&&\times P(n_A,n_I+1,t|{\cal S};{\cal I}_{n_A',n_I'})\nonumber\\
&&\times P({\cal S}|{\cal I}_{n_A',n_I'})\nonumber\\
&&\times P({\cal I}_{n_A',n_I'})\label{cond1}.
\end{eqnarray}
Now, it is $P(n_A,n_I,t+\Delta t|n_A,n_I+1,t;{\cal S};{\cal I}_{n_A',n_I'})=1$, as the particle that was infected in $(t-\tau,t-\tau+\Delta t)$ and survived up to $t$, dies with probability $1$ during $(t,t+\Delta t)$, according to the rules of the process. The  conditional probability of the second line of the rhs is the probability that there are $n_A,n_I+1$ particles at time $t$ given that there were $n_A',n_I'$ particles at time $t-\tau$ and one $X_A$ particle was infected in $(t-\tau,t-\tau+\Delta t)$ and survived up to time $t$. Since the presence of the infected particle does not influence the dynamics of births, deaths and infection of other particles (remember that we have assumed that $X_I$ are ``inert" particles such that the different rates do not depend on the number of $X_I$ particles), this conditional probability can be simply obtained by considering the process in which the dynamics of the infected particle is decoupled from the rest of the particles, i.e. as if removing that particle from the dynamical process. This leads  to:
\begin{equation}
 P(n_A,n_I+1,t|{\cal S};{\cal I}_{n_A',n_I'})=P(n_A,n_I,t|n_A'-1,n_I',t-\tau+\Delta t)\label{trick}.
\end{equation}
 Furthermore, as all $X_I$ present at $t-\tau$ have died before $t$ (except the one infected during $(t-\tau,t-\tau+\Delta t$)), this conditional probability is, in fact, independent of $n'_I$ and, for translational invariance, depends only on the time difference $\tau$, so we can write it as $P(n_A,n_I,t|n_A'-1,t-\tau)$. 
The next term, $P({\cal S}|{\cal I}_{n_A',n_I'})$,  corresponds to the survival probability during a time interval of length $\tau$, or $e^{-\gamma'\tau}$. Finally, $P({\cal I}_{n_A',n_I'})$ is the infection probability ${\beta}n'_A\Delta tP(n_A',n_I',t-\tau)$. Collecting all the terms, the contribution to the rhs of Eq.(\ref{generaldelayme}) is $\sum_{n_A'}P(n_A,n_I,t|n_A'-1,t-\tau+\Delta t){\beta}n'_AP(n_A',t-\tau)e^{-\gamma'\tau}\Delta t$.
\\
(5) None of the above processes occur in the interval $(t,t+\Delta t)$: $(n_A,n_I)\to (n_A,n_I)$. 
\small
\begin{eqnarray}
&P&(n_A,n_I,t+\Delta t;n_A,n_I,t)=P(n_A,n_I,t+\Delta t|n_A,n_I,t)P(n_A,n_I,t)\\
&&=\left[1-C\Delta t-(\gamma+{\beta})n_A\Delta t-\gamma'n_I\Delta t-\sum_{n'_A,n'_I}P(n_A,n_I,t+\Delta t;{\cal S};{\cal I}_{n_A',n_I'}|n_A,n_I,t)\right]P(n_A,n_I,t)\nonumber\\
&&=\left[1-C\Delta t-(\gamma+{\beta})n_A\Delta t-\gamma'n_I\Delta t\right]P(n_A,n_I,t)-\sum_{n'_A,n'_I}P(n_A,n_I,t+\Delta t;n_A,n_I,t;{\cal S};{\cal I}_{n_A',n_I'}).\nonumber
\end{eqnarray}
\normalsize
Treating the last term in the same way  as in the previous case  (4b), one obtains that the contribution to the rhs of Eq.(\ref{generaldelayme}) is:
\begin{eqnarray}
P(n_A,n_I,t&+&\Delta t;n_A,n_I,t)=\left[1-C\Delta t-(\gamma+{\beta})n_A\Delta t-\gamma'n_I\Delta t\right]P(n_A,n_I,t)\nonumber\\
&&-\sum_{n_A'}P(n_A,n_I-1,t|n_A'-1,t-\tau){\beta}n'_AP(n_A',t-\tau)e^{-\gamma'\tau}\Delta t.
\end{eqnarray}
After adding up all contributions to the rhs of  Eq.(\ref{generaldelayme}), dividing by $\Delta t$ and taking the limit $\Delta t\to 0$ we obtain the master equation of the process:
\begin{widetext}
\begin{eqnarray}
\label{me}
\frac{dP(n_A,n_I,t)}{dt}&=&(E_A^{-1}-1)CP(n_A,n_I,t)+(E_A-1)\gamma n_AP(n_A,n_I,t)\nonumber\\
&&+(E_AE_I^{-1}-1){\beta}n_AP(n_A,n_I,t)+(E_I-1)\gamma'n_IP(n_A,n_I,t)\nonumber\\
&&+(E_I-1)\sum_{n_A'=0}^{\infty}P(n_A,n_I-1,t|n_A'-1,t-\tau){\beta}n_A'P(n_A',t-\tau)e^{-\gamma'\tau},
\end{eqnarray}
\end{widetext}
where $E_{A,I}$ are step operators, $E^k_{A,I}f(n_{A,I})=f(n_{A,I}+k)$ and $P(n_A,t)=\sum_{n_I}P(n_A,n_I,t)$ is the marginal probability.
As written, this is an equation for the one-time probability, and has to be supplemented with the appropriate initial conditions. The evolution equation for the conditional probability $P(n_A,n_I,t|n_A^0,n_I^0,t_0)$ will be considered in section \ref{correlations}. For Markovian processes this equation would be the same as Eq.(\ref{me}) conditioning all appearing probabilities to $(n_A^0,n_I^0,t_0)$, but for non-Markovian processes this is not necessarily the case.
 
The next step is to calculate the conditional probability appearing in Eq.(\ref{me}). Let $P^*(n_A,n_I,t|n_A^0,n_I^0,0)$ be the probability that there are $n_A$ particles of $X_A$ type and $n_I$ particles of $X_I$ type at time $t$ given the initial condition that there were $n_A^0$ and $n_I^0$ particles at time $t=0$ and {\bf under the condition that particles can not die through delayed-degradation}. The key point is to note that:
\begin{equation}
 P(n_A,n_I,t|n_A',t-\tau)=P^*(n_A,n_I,\tau|n_A',0,0).
\end{equation}
This is so because, as reminded before, all $X_I$-particles present at any time die before a interval of length $\tau$ passes and their presence does not influence the dynamics of births, deaths and infections of other particles. Note that this equality is only valid as written, as in general $P(n_A,n_I,t|n_A',t-t')\neq P^*(n_A,n_I,t'|n_A',0,0)$, the equality only occurring for $t'=\tau$.  For $t'<\tau$ not all the infected particles present at $t-t'$ have died at $t$; for $t'>\tau$ some of the particles infected after $t-t'$ may have died trough delay-degradation before $t$. Therefore, we need to compute $P^*(n_A,n_I,t|n_A'-1,0,0)$ (we simplify notation to $P^*(n_A,n_I,t|n_A'-1)$). It follows the same master equation (\ref{me}) without the last term. For the sake of completeness, we write down the final system of equations:
\begin{widetext}
\begin{eqnarray}
\label{me2}
\frac{dP(n_A,n_I,t)}{dt}&=&(E_A^{-1}-1)CP(n_A,n_I,t)+(E_A-1)\gamma n_AP(n_A,n_I,t)\nonumber\\
&&+(E_AE_I^{-1}-1){\beta}n_AP(n_A,n_I,t)+(E_I-1)\gamma'n_IP(n_A,n_I,t)\nonumber\\
&&+(E_I-1)\sum_{n_A'=0}^{\infty}P^*(n_A,n_I-1,\tau|n_A'-1){\beta}n_A'P(n_A',t-\tau)e^{-\gamma'\tau},
\end{eqnarray}
\end{widetext}
\small
\begin{widetext}
\begin{eqnarray}
\label{me3}
\frac{dP^*(n_A,n_I,t|n_A'-1)}{dt}&=&(E_A^{-1}-1)CP^*(n_A,n_I,t|n_A'-1)+(E_A-1)\gamma n_AP^*(n_A,n_I,t|n_A'-1)\nonumber\\
&&+(E_AE_I^{-1}-1){\beta}n_AP^*(n_A,n_I,t|n_A'-1)\nonumber\\
&&+(E_I-1)\gamma'n_IP^*(n_A,n_I,t|n_A'-1)
\end{eqnarray}
\end{widetext}
\normalsize

The process defined by Eq.(\ref{me3}) is a standard Markovian process for which exact and approximate schemes have been developed to find its solution. Once we have obtained $P^*(n_A,n_I,\tau|n_A'-1)$ by solving Eq.(\ref{me3}) with the appropriate initial condition  $P^*(n_A,n_I,0|n_A'-1)=\delta_{n_A,n_A'-1}\delta_{n_I,0}$,  we can replace in Eq.(\ref{me2}) and proceed to solve for $P(n_A,n_I,t)$. A very convenient procedure is to derive a differential equation for the generating function $G(s_A,s_I,t)\equiv\sum_{n_A,n_I}s_A^{n_A}s_I^{n_I}P(n_A,n_I,t)$ in terms of the corresponding generating function of the process without delay-degradation, $G^*(s_A,s_I,t)$, see the Appendix for details. Note that the active variables $X_A$ follow an independent dynamics as can be seen by the rules of the process (the presence of $X_I$ molecules does not alter at all the dynamics of $X_A$) or, from a more formal point of view, by noticing that, as derived from Eq.(\ref{me2}), the marginal probabilities for the numbers $n_A$, $P(n_A,t)$ follow a closed master equation of a birth-death process:
\begin{widetext}
\begin{eqnarray}
\label{me4}
&&\frac{dP(n_A,t)}{dt}=(E_A^{-1}-1)CP(n_A,t)+(E_A-1)\gamma n_AP(n_A,t).
\end{eqnarray}
\end{widetext}

\section{Solution in the case of constant rates}\label{exactrates}
Up to now, we have been rather general. The only assumption needed is that the rates can depend only on the number of active particles $n_A$. We now provide the exact solution in the case of constant rates.  In this case $P^*$ corresponds to a simple birth-death Poisson process. As shown in the Appendix, it turns out that if the initial condition $P(n_A,n_I,0)$ follows a Poisson distribution (this includes the case which starts with no molecules at all at $t=0$) then the generating function at arbitrary time is  $G(s_A,s_I,t)=e^{x_{A}(t)(s_A-1)+x_I(t)(s_I-1)}$. This shows that the joint probability of $n_A$ and $n_I$ is the product of independent Poisson distributions, with mean values and variances
$\langle n_A(t)\rangle=\sigma^2_{n_A}(t)=x_A(t)$,  $\langle n_I(t)\rangle=\sigma^2_{n_I}(t)=x_I(t)$
. It follows that the total number of $X$ particles, $n=n_A+n_I$, also obeys at all times a Poisson distribution with parameter $x(t)=x_A(t)+x_I(t)$. If we start with an initial condition different from Poisson-distributed, the time-dependent solution is not Poissonian, but this form is recovered in the steady state with mean values $\langle n_{A,I}\rangle_{st}$. The mean values satisfy differential equations including delay terms:
\begin{eqnarray}
 \frac{dx_A(t)}{dt}&=&C-ax_A(t),\label{average}\label{eqni} \\
 \frac{dx_I(t)}{dt}&=&-\gamma'x_I(t) +{\beta}(x_A(t) -e^{-\gamma'\tau}x_A(t-\tau) )\nonumber,
\end{eqnarray}
with $a\equiv {\beta}+\gamma$. The solution with initial condition $x_A(t\le 0)=0$, $x_I(t=0)=0$ is:
\begin{eqnarray}
\label{eq:solxa}
x_A(t)&&=\frac{C}{a} \left(1-e^{-a t}\right), \\
x_I(t)&&=
\begin{cases}\frac{C {\beta}}{a-\gamma'}\left[\frac{1-e^{-\gamma' t}}{\gamma'}-\frac{1-e^{-a t}}{a}\right],&0\le t\le\tau,\\
\frac{C{\beta}}{a}\left[\frac{1-e^{-\gamma'  \tau }}{\gamma'}+\frac{(1-e^{\tau  (a-\gamma')})}{a-\gamma'} e^{-at }\right],&t\ge\tau.
\end{cases}\nonumber
\end{eqnarray}
The steady-state values are $\langle n_A\rangle_{st}=C/a$, $\langle n_I\rangle_{st}=C{\beta}(1-e^{-\gamma'\tau})/(a\gamma')$. 

It is important to note that it is not possible to write a closed equation for the evolution of the total number of particles $x(t)=x_A(t)+x_I(t)$. In the case $\gamma=\gamma'$ we arrive at $\dot x(t)=C-\gamma x(t)-{\beta}e^{-\gamma'\tau}x_A(t-\tau)$, and in the case $\gamma'=0$ we get $\dot x(t)=C-\gamma x_A(t)-{\beta}x_A(t-\tau)$ which differs in both cases from the closed result $\dot x(t)=C-\gamma x(t)-{\beta}x(t-\tau)$ used in \cite{bratsun}. 
A different process that leads to this closed form of the rate equations corresponds to delayed production and linear negative feedback:
\begin{equation}
0 {{C-{\beta}n \atop \Longrightarrow}\atop{\tau}} X, \,X {{\gamma \atop \longrightarrow}\atop{}} 0.
\end{equation}
However, in this case the creation rate $C-{\beta}n$ may become negative, and so the process is ill-defined \cite{creationdelay}. 

\section{Time correlations}\label{correlations}
We now turn to the calculation of the time correlations. For this, we will write a master equation for the conditional probability $P(n_A,n_I,t|n_A^0,n_I^0,t_0)$ and derive evolution equations for the conditional averages $\langle n_{A,I},t|n_A^0,n_I^0,t_0\rangle$. As commented before, for a Markovian process, $P(n,t)$ and $P(n,t|n^0,t_0)$ satisfy identical master equations but with different initial conditions. For non-Markovian processes, however, in principle one has to specify the whole history before $t_0$. In our system, the problem is that the dynamics from $t_0$ depends on when the $n_I^0$ particles were created. We are mainly interested in the steady state. The term that causes difficulties is the one corresponding to delayed-degradation. To obtain the proper contribution to the master equation, and in the line of the previous arguments, we take a starting point similar to Eq.(\ref{cond1}):
\begin{eqnarray}\label{cond}
P(n_A,n_I,t+\Delta t;n_A,n_I+1,t;{\cal S};{\cal I}|n_A^0,n_I^0,t_0)&=&1\times P(n_A,n_I+1,t|{\cal S};{\cal I};n_A^0,n_I^0,t_0)\nonumber\\
&&\times P({\cal S};{\cal I}|n_A^0,n_I^0,t_0),
\end{eqnarray}
with ${\cal I}=\bigcup_{n_A',n_I'}{\cal I}_{n_A',n_I'}$ the event in which a particle was infected during the interval $(t-\tau,t-\tau+\Delta t)$, independently of the number of particles present at that time.

One has to distinguish the intervals $t-t_0<\tau$ and $t-t_0>\tau$. In the former,
following the reasoning of Eq.(\ref{trick}) we see that the first conditional probability in the rhs of Eq.(\ref{cond}) equals $P(n_A,n_I,t|n_A^0,n_I^0-1,t_0)$, since, as commented before, one of the $X_I$ particles present at time $t_0$ will stay until time $t$, and its presence does not influence dynamics of creations and deaths of other particles. Next, we use Bayes' theorem:
\begin{equation}
P({\cal S};{\cal I}|n_A^0,n_I^0,t_0)=\sum_{n_A',n_I'}P({\cal S};{\cal I}_{n_A',n_I'}|n_A^0,n_I^0,t_0)=\sum_{n_A',n_I'}P(n_A^0,n_I^0,t_0|{\cal S};{\cal I}_{n_A',n_I'})\frac{P({\cal S};{\cal I}_{n_A',n_I'})}{P(n_A^0,n_I^0,t_0)}.
\end{equation}
An argument similar to the one used in deriving Eq.(\ref{trick}) shows that $
P(n_A^0,n_I^0,t_0|{\cal S};{\cal I}_{n_A',n_I'})=P(n_A^0,n_I^0-1,t_0|n_A'-1,n_I',t-\tau)$ which can be written as $P(n_A'-1,n_I',t-\tau|n_A^0,n_I^0-1,t_0)P(n_A^0,n_I^0-1,t_0)/P(n_A'-1,n_I',t-\tau)$. We use also that $P({\cal S};{\cal I}_{n_A',n_I'})=n_A'P(n_A',n_I',t-\tau)e^{-\gamma'\tau}{\beta}\Delta t $ to obtain: 
\begin{eqnarray}
P({\cal S};{\cal I}|n_A^0,n_I^0,t_0)&=&\sum_{n_A',n_I'}P(n_A'-1,n_I',t-\tau|n_A^0,n_I^0-1,t_0)n_A'{\beta}e^{-\gamma'\tau}\Delta t
\frac{P(n_A',n_I',t-\tau)}{P(n_A'-1,n_I',t-\tau)}\nonumber\\
&&\cdot\frac{P(n_A^0,n_I^0-1,t_0)}{P(n_A^0,n_I^0,t_0)}.
\end{eqnarray}
We know that in the steady state, if the rates are constant, $n_A$ and $n_I$ follow independent Poisson distributions. This allows us to compute the ratios of probabilities in this expression:
\begin{eqnarray}
\frac{P(n_A',n_I',t-\tau)}{P(n_A'-1,n_I',t-\tau)}&=&\frac{\langle n_A\rangle_{st}}{n_A},\\
\frac{P(n_A^0,n_I^0-1,t_0)}{P(n_A^0,n_I^0,t_0)}&=&\frac{n_I^0}{\langle n_I\rangle_{st}}.
\end{eqnarray}
Which leads to:
\begin{eqnarray}
P({\cal S};{\cal I}|n_A^0,n_I^0,t_0)&=&\sum_{n_A',n_I'}P(n_A'-1,n_I',t-\tau|n_A^0,n_I^0-1,t_0){\beta}e^{-\gamma'\tau}\Delta t\frac{\langle n_A\rangle_{st}}{\langle n_I\rangle_{st}}n_I^0\\
&=&{\beta}e^{-\gamma'\tau}\Delta t\frac{\langle n_A\rangle_{st}}{\langle n_I\rangle_{st}}n_I^0,
\end{eqnarray}
where we have used $\sum_{n_A',n_I'}P(n_A'-1,n_I',t-\tau|n_A^0,n_I^0-1,t_0)=1$.
Putting all the pieces together we find that the delay degradation term in the master equation for the conditional probability for $t-t_0<\tau$ is:
 \begin{equation}
 (E_I-1)P(n_A,n_I-1,t|n_A^0,n_I^0-1,t_0)\frac{\gamma'}{e^{\gamma'\tau}-1}n_I^0\label{conditionme}.
\end{equation}

On the other hand, for $t-t_0>\tau$, expression (\ref{cond}) equals:
\begin{eqnarray}
\sum_{n_A',n_I'}P(n_A,n_I,t|n_{A}'-1,n_{I}',t-\tau;n_A^0,n_I^0,t_0)e^{-\gamma'\tau}n_A'{\beta}\Delta tP(n_A',n_I',t-\tau|n_A^0,n_I^0,t_0)\label{conditionme2},
\end{eqnarray}
which is the same as the delay-degradation term in the master equation Eq.(\ref{me}) but conditioning all probabilities to $n_A^0,n_I^0$ at time $t_0$, as happens in Markovian processes.

Using (\ref{conditionme}) and (\ref{conditionme2}) in the corresponding time intervals, we obtain the evolution of the conditional averages in the steady state:
\footnotesize
\begin{widetext}
\begin{eqnarray}
\frac{d\langle n_A,t|n_A^0,n_I^0,t_0\rangle}{dt}&=&C-a\langle n_A,t|n_A^0,n_I^0,t_0\rangle.\\
 \frac{d\langle n_I,t|n_A^0,n_I^0,t_0\rangle}{dt}&=&
\begin{cases}-\gamma\langle n_I,t|n_A^0,n_I^0,t_0\rangle+{\beta}\langle n_A,t|n_A^0,n_I^0,t_0\rangle-\frac{\gamma'}{e^{\gamma'\tau}-1}n_I^0,&t_0\le t\le t_0+\tau,\\
 -\gamma\langle n_I,t|n_A^0,n_I^0,t_0\rangle+{\beta}(\langle n_A,t|n_A^0,n_I^0,t_0\rangle-e^{-\gamma'\tau}\langle n_A,t-\tau|n_A^0,n_I^0,t_0\rangle),&t\ge t_0+\tau\label{meaneq}.
\end{cases}\nonumber
\end{eqnarray}
\end{widetext}
\normalsize

The steady state correlations, \makebox{$K_{UV}(t)\equiv\langle n_{U}(t'+t)n_{V}(t')\rangle_{st}-\langle n_{U}\rangle_{st}\langle n_{V}\rangle_{st}$} can be obtained integrating the previous equations and averaging over initial conditions in the steady state.  The result is:
\begin{eqnarray}
 K_{AA}(t)&=&\langle n_{A}\rangle_{st}e^{-at},\\
K_{II}(t)&=&
\begin{cases}
\langle n_I\rangle_{st} \frac{e^{-\gamma't}-e^{-\gamma'\tau}}{1-e^{-\gamma'\tau}},&0\le t\le\tau,\label{ki}\\
0,&t\ge\tau,
\end{cases}\\
K_{IA}(t)&=&
\begin{cases}
\frac{C{\beta}}{a}\frac{e^{-\gamma't}-e^{-at}}{a-\gamma'},&0\le t\le\tau,\\
\frac{C{\beta}}{a}\frac{e^{(a-\gamma')\tau}-1}{a-\gamma'}e^{-at},&t\ge\tau,     
\end{cases}\\
K_{AI}(t)&=&0.
\end{eqnarray}
The autocorrelation function for the total number of particles $n=n_A+n_I$ is $K(t)=K_{AA}(t)+K_{II}(t)+K_{IA}(t)+K_{AI}(t)$. Note that in the case $\gamma'=0$ the correlation for the $X_I$ particles decays linearly in time instead of the typical exponential decay (as can be seen taking the limit $\gamma'\rightarrow0$ in Eq.(\ref{ki})). For arbitrary $\gamma'$ the correlation function for the $X_I$ particles drops strictly to zero at $t=\tau$.

An analysis of the previous expressions shows that, in all cases, the autocorrelation functions $K_{AA}(t)$, $K_{II}(t)$, and $K(t)$ decrease monotonically in time, so there is no signature of stochastic oscillations, as found in the particular case, $\gamma'=0$, considered in \cite{Mieckisz}. These exact theoretical expressions are in excellent agreement with the results of extensive numerical simulations of the stochastic process (\ref{stoproc}) using a conveniently modified form of the Gillespie's algorithm\cite{gillespie,cai}. This agreement provides an independent check of the correctness of the theoretical calculations. An important ingredient of the process is that particles which are infected at, say, time $t$ and are then scheduled to die at time $t+\tau$, can nevertheless die instantaneously at a rate $\gamma'$ during the interval $(t,t+\tau)$. If this is the case, in the stochastic simulation one has to remove that particle from the list of scheduled events to happen at $t+\tau$. Otherwise, one is removing a particle that was already removed. Stochastic oscillations do appear if one, erroneously, removes twice these particles.

\section{Discussion}\label{conclusions}
In this work, we have studied a stochastic model of protein level dynamics that includes delay in the degradation. The exact solution shows that no oscillations are present in the system, contrary to previous results, and in agreement with a recent analysis in a simplified version of the stochastic model that allows for a Markovian reduction \cite{Mieckisz}. This implies that the presence of delay in degradation alone cannot give rise to oscillations. We have also analyzed the derivation of master equations and the time correlations in systems with delay, pointing out the differences with Markovian processes. Our approach allows one to deal with more general systems that the one used in \cite{Mieckisz} and may be of interest to study other non-Markovian processes. 
Exact results are specially valuable in small systems where approximated schemes typically fail. They are also important to study the validity of assumptions and approximations to be used in more complicated systems.

In a broader context, due to the important role of stochasticity and delay in many biological, physical or technological systems, the present work seems relevant as a case where exact results can be obtained, allowing to clarify the combined effect of stochasticity and delay.

\section{Appendix: Calculation of the probabilities}\label{appendix}
Let us define the generating function of $P^*(n_A,n_I,t|n_A^0,n_I^0,0)$ as $G^*(s_A,s_I,t)\equiv\displaystyle\sum_{n_A=0}^{\infty}\sum_{n_I=0}^{\infty}s_A^{n_A}s_I^{n_I}P^*(n_A,n_I,t|n_A^0,n_I^0,0)$. Using standard techniques, it follows from the master equation (\ref{me3})  that it obeys the partial differential equation:
\begin{eqnarray}
 \frac{\partial G^*}{\partial t}=[\gamma(1-s_A)+{\beta}(s_I-s_A)]\frac{\partial G^*}{\partial s_A}+\gamma'(1-s_I)\frac{\partial G^*}{\partial s_I}+C(s_A-1)G^*
 \end{eqnarray}
 with initial condition $G^*(s_A,s_I,0)=s_A^{n_A^0}s_I^{n_I^0}$. 
By the method of the characteristics, one can find that the solution in the case of the initial condition $n_A^0=n'_A-1,n_I^0=0$ (as needed in the master equation Eq.(\ref{me})) can be written as:
\begin{equation}
\label{eq:solg}
G^*(s_A,s_I,t)=\left[1+\Phi(s_A,s_I,t)\right]^{n'_A-1}\exp\left[C\int_0^tdt'\,\Phi(s_A,s_I,t')\right],
\end{equation}
where
\begin{equation}
\Phi(s_A,s_I,t)=(s_A-1){\rm e}^{-at}+(s_I-1)\frac{{\beta}}{a-\gamma'}\left({\rm e}^{-\gamma't}-{\rm e}^{-at}\right).
\end{equation}

It follows from Eq.(\ref{me2}) that the generating function for the original delayed process $G(s_A,s_I,t)\equiv\displaystyle\sum_{n_A=0}^{\infty}\sum_{n_I=0}^{\infty}s_A^{n_A}s_I^{n_I}P(n_A,n_I,t)$ obeys:
\begin{eqnarray}
 \frac{\partial G}{\partial t}&=&[\gamma(1-s_A)+{\beta}(s_I-s_A)]\frac{\partial G}{\partial s_A}+\gamma'(1-s_I)\frac{\partial G}{\partial s_I}+C(s_A-1)G\label{generating}\\
&+&{\beta}e^{-\gamma'\tau}(1-s_I)\sum_{n'_A=0}G^*(s_A,s_I,\tau)n'_AP(n'_A,t-\tau)\nonumber,
\end{eqnarray}
We assume $n_A(t)=n_I(t)=0$ for $t<0$ so that both the master equation (\ref{me}) and the equation for the generating function (\ref{generating}) are valid for $t>0$.
The number of $X_A$ particles follows a Poisson distribution at all times $P(n_A,t)=\frac{x_A(t)^{n_A}}{n_A!}e^{-x_A(t)(s_A-1)}$ with $x_A(t)$ the average value given in Eq.(\ref{eq:solxa}) (assuming the initial condition is $n_A(0)=0$). Using Eq.(\ref{eq:solg}), it is possible to perform the sum over the $n_A'$ variable to obtain the equation:
\begin{eqnarray}
 \frac{\partial G}{\partial t}&=&[\gamma(1-s_A)+{\beta}(s_I-s_A)]\frac{\partial G}{\partial s_A}+\gamma'(1-s_I)\frac{\partial G}{\partial s_I}\nonumber\\
&+&C(s_A-1)G+{\beta}e^{-\gamma'\tau}(1-s_I)x_A(t-\tau)\\
&\times&\exp\left[x_A(t-\tau)\Phi(s_A,s_I,\tau)+C\int_0^{\tau}dt'\,\Phi(s_A,s_I,t')\right]\nonumber.
\end{eqnarray}

One can check by direct substitution that the Poisson distribution $G(s_A,s_I,t)=e^{x_{A}(t)(s_A-1)+x_I(t)(s_I-1)}$ is a solution of this equation if $x_A(t)$ and $x_I(t)$ obey the differential equations (\ref{eqni}). Therefore, $n_A$ and $n_I$ follow independent Poisson distributions at all times (assuming $n_A(0)=n_I(0)=0$). It follows that the total number of $X$ particles, $n=n_A+n_I$, also obeys at all times a Poisson distribution with parameter $x(t)=x_A(t)+x_I(t)$. If we start with an initial condition different from $n_A(0)=n_I(0)=0$ (or $n_A$ Poisson-distributed), the time-dependent solution is not Poissonian, but this form is recovered in the steady state.

{\textbf{Acknowledgments:}}
We thank  E.A. Herrada and J. Garcia-Ojalvo for useful discussions and L.G. Morelli for pointing out references \cite{bodnar,Mieckisz} to us. We acknowledge financial support by the MEC (Spain) and FEDER (EU) through project FIS2007-60327. L.F.L. is supported by the JAEPredoc program of CSIC.

\end{document}